\documentclass[aps,rmp,eqsecnum,amssymb,12pt]{article}
\usepackage{amssymb}
\usepackage{amsmath}
\usepackage{epsfig}
\usepackage{natbib}
\usepackage[dvips]{color}

\def\beq{\begin{equation}}
\def\eeq{\end{equation}}

\def\beqa{\begin{eqnarray}}
\def\eeqa{\end{eqnarray}}

\begin{document}

\begin{titlepage}

\setcounter{page}{1} \baselineskip=15.5pt \thispagestyle{empty}

\begin{flushright}
PUPT-2221\\
hep-th/0701064\\
\end{flushright}
\vfil

\begin{center}
{\LARGE Gauge/Gravity Duality and Warped Resolved Conifold}
\end{center}
\bigskip\

\begin{center}
{\large Igor R. Klebanov and Arvind Murugan}
\end{center}

\begin{center}
\textit{Department of Physics and Princeton Center for Theoretical
Physics\\ Princeton University, Princeton, NJ 08544}
\end{center} \vfil

\noindent We study supergravity backgrounds encoded through the
gauge/string correspondence by the $SU(N) \times SU(N)$ theory
arising on $N$ D3-branes on the conifold. As discussed in
hep-th/9905104, the dynamics of this theory describes warped
versions of both the singular and the resolved conifolds through
different (symmetry breaking) vacua. We construct these
supergravity solutions explicitly and match them
with the gauge theory with different sets of vacuum expectation
values of the bi-fundamental fields $A_1, A_2, B_1, B_2$. For the
resolved conifold, we find a non-singular $SU(2)\times U(1)\times
U(1)$ symmetric warped solution produced by a stack of D3-branes
localized at a point on the blown-up 2-sphere. It describes a
smooth RG flow from $AdS_5 \times T^{1,1}$ in the UV to $AdS_5
\times S^5$ in the IR, produced by giving a VEV to just one field,
e.g. $B_2$. The presence of a condensate of 
baryonic operator ${\rm det} B_2$ is confirmed using a Euclidean D3-brane
wrapping a 4-cycle inside the resolved conifold. 
The Green's functions on the singular and resolved
conifolds are central to our calculations and
are discussed in some detail.

\vfil

\end{titlepage}

%=========================================================================
\newpage

\tableofcontents

\pagestyle{headings}

\section{Introduction}
\label{sec-intro} The basic AdS/CFT correspondence
\cite{Maldacena,GKP,Witten} (see
\cite{Aharony:1999ti,Klebanov:2000me} for reviews) is motivated by
considering the low energy physics of a heavy stack of D3-branes at
a point in flat spacetime. Taking the near-horizon limit of this
geometry motivates a duality between type IIB string theory on
$AdS_5 \times S^5$ and ${\cal N}=4$ $SU(N)$ supersymmetric
Yang-Mills gauge theory. This correspondence was generalized
to theories with ${\cal N}=1$ superconformal symmetry in
\cite{KW,Morrison} by considering a stack D3-branes, not in flat
space, but placed at the tip of a 6d Calabi-Yau cone $X_6$. The near
horizon limit in this case turns out to be $AdS_5\times Y_5$ where
$Y_5$ is the compact 5 dimensional base of $X_6$ and is a
Sasaki-Einstein space.

Among the simplest of these examples is $Y_5 = T^{1,1}$, corresponding
$X_6$ being the conifold. It was found that the low-energy
gauge theory on the D3-branes at the tip of the conifold is a ${\cal
N}=1$ supersymmetric $SU(N)\times SU(N)$ gauge theory with
bi-fundamental chiral superfields $A_i$, $B_j$ $(i,j=1,2)$ in
$(N,\bar{N})$ and $(\bar{N},N)$ representations of the gauge groups,
respectively \cite{KW,Morrison}. The superpotential for this gauge
theory is $W \sim {\rm Tr} \det A_i B_j = {\rm Tr}\ (A_1 B_1 A_2 B_2 -
A_1 B_2 A_2 B_1)$. The continuous global symmetries of this theory
are $SU(2) \times SU(2) \times U(1)_R \times U(1)_B$ where the $SU(2)$
factors act on
$A_i$ and $B_j$ respectively, $U(1)_B$ is a baryonic
symmetry, and $U(1)_R$ is the R-symmetry with $R_A=R_B=\frac{1}{2}$.
This assignment ensures that $W$ is marginal, and one can also show
that the gauge couplings do not run. Hence this theory is
superconformal for all values of gauge couplings and superpotential
coupling \cite{KW,Morrison}.

  When the above gauge theory is considered with no vacuum expectation values
  (VEV's) for any of the fields,
  we have a superconformal theory with the $AdS_5\times T^{1,1}$ dual. In \cite{KW2},
  more general vacua of this theory were studied. It was argued that moving the D3-branes off the tip
  of the singular conifold corresponds to a symmetry breaking in the gauge theory due to VEV's for
  the $A,B$ matter fields such that the VEV of operator
\begin{equation}
  {\cal U} = {1\over N}
{\rm Tr} \left (|B_1|^2 +|B_2|^2-|A_1|^2 -|A_2|^2 \right )
  \end{equation}
  vanishes.
  Further, more general vacua exist for this theory
  in which this operator acquires a non-zero VEV.\footnote{
  As was pointed out in \cite{KW}, no D-term equation
constrains this
  operator since the $U(1)$ gauge groups decouple in the infrared.}
  It was pointed out in \cite{KW2} that these vacua cannot correspond to D3-branes on the singular
  conifold.
  Instead, such vacua with ${\cal U} \neq 0$ correspond to D3-branes on the resolved conifold.
This ``small resolution'' is a motion along the K\" ahler moduli space
  where the singularity of the conifold
is replaced by a finite $S^2$. Thus the $SU(N)\times SU(N)$ gauge theory was argued to
  incorporate in its different vacua both the singular and resolved conifolds.
  On the other hand, the deformation of the conifold, which is a motion along
the complex structure moduli space, can be achieved through
replacing
  the gauge theory by the cascading $SU(N) \times SU(N+M)$ gauge theory (see \cite{KS}).

One of the goals of this paper is to construct the warped SUGRA
solutions corresponding to the gauge theory vacua with
  ${\cal U} \neq 0$. Our work builds on the earlier resolved conifold solutions
  constructed by Pando Zayas and Tseytlin
  \cite{PandoZayas}, where additional simplifying symmetries were sometimes imposed.
  Such solutions corresponding to D3-branes ``smeared'' over a
  region were found to be singular in the IR \cite{PandoZayas}.
We will instead look for ``localized'' solutions
  corresponding to the whole D3-brane stack located at one point on the (resolved) conifold.
  This corresponds to giving VEV's to the fields $A_i, B_j$ which are proportional
to $1_{N\times N}$.  We construct the
duals of these gauge theory vacua and find them
to be completely non-singular. The solution acquires a particularly simple
form when the stack is placed at the north pole of the blown up
2-sphere at the bottom of the resolved conifold.
  It corresponds to the simplest way to have ${\cal U} \neq 0 $ by setting
  $B_2= u 1_{N\times N}$ while
  keeping $A_1 = A_2 = B_1 =0 $.

 Following \cite{Balasubramanian,KW2}, we also interpret our solutions
%this construction in the gauge theory
 as having
 an infinite series of VEV's for various operators in addition to
${\cal U}$. For this, we rely on the relation between
 normalizable SUGRA modes and gauge theory VEV's in the AdS/CFT dictionary.
 When a given asymptotically AdS
solution has a (linearized) perturbation that falls off as $r^{-\Delta}$ at large $r$,
 it corresponds to assigning a VEV for a certain operator ${\cal O}$ of dimension $\Delta$ in the dual
 gauge theory \cite{Balasubramanian,KW2}.
 The warp factor produced by a stack of D3-branes on the resolved conifold is related to
 the Green's function on
 the resolved conifold. This warp factor can be expanded in harmonics and corresponds to a series of
 normalizable fluctuations as above, and hence a series of operators in the gauge theory
 acquire VEV's.\footnote{In the ${\cal N}=4$ SUSY example,
the normalizations of the VEV's have been matched with the size of
the SUGRA perturbations around $AdS_5\times S^5$
 (see \cite{deHaro,Skenderis1,Skenderis2}).
 In this paper we limit ourselves to a more qualitative picture where the
precise normalizations
 of the VEV's are not calculated.}
 For this purpose, we write the harmonics in a convenient set of variables $a_i,b_j$ that makes the
 link with gauge theory operators built from $A_i,B_j$
 immediate.
Due to these symmetry breaking VEV's, the gauge theory flows from the $SU(N)\times
SU(N)$ ${\cal N}=1$ theory in
   the UV to the $SU(N)$ ${\cal N}=4$ theory in the IR,
as one would expect when D3-branes are placed at a
   smooth point.  The SUGRA solution is shown to have two asymptotic AdS regions --
an $AdS_5 \times T^{1,1}$ region in the UV, and also an $AdS_5   \times S^5$ region
   produced in the IR by the localized stack of D3-branes.
   This can be considered an example of holographic RG flow.
The Green's functions determined here might also have applications to models of
   D-brane inflation,
   and to computing 1-loop corrections to gauge couplings in gauge theories
   living on cycles in the geometry \cite{Giddings,BDKMMM}.

   When the branes are placed on the blown up 2-sphere at the bottom of the resolved
conifold, this corresponds to
$A_1= A_2=0$ in the gauge theory.
Hence no chiral mesonic operators, such as ${\rm Tr} A_i B_j$, have VEV's,
but baryonic operators, such as $\det B_2$, do acquire VEV's. Therefore,
such solutions, parametrized by the size of the resolution and position of the stack on the 2-sphere,
are dual to a ``non-mesonic'' (or ``baryonic'') branch of the $SU(N)\times SU(N)$ SCFT
(see \cite{Benvenuti} for a related discussion). These solutions have a blown up $S^2$.
On the other hand, the
solutions dual to the baryonic branch of the cascading $SU(N) \times SU(N+M)$
gauge theory were constructed in
\cite{Butti,Dymarsky} (for an earlier linearized treatment, see \cite{GHK}) and have a blown up
$S^3$ supported by the 3-form flux.

The paper is organized as follows.
 In Section \ref{sec-conifold}, we review and establish notation for describing the conifold,
 its resolution, its symmetries and coordinates that make the symmetries manifest.
We also review 
the metric of the resolved conifold and the
 singular smeared solution found in \cite{PandoZayas}.
 In Section \ref{sec-singular}, as a warm up, we study the simple example of moving a stack of D3-branes
 away from the tip of the singular conifold. We present the explicit supergravity solution for this
 configuration by determining the Green's function on the conifold. We interpret the operators that get
 VEV's and note that in general, chiral as well as non-chiral operators get VEV's.
 In Section \ref{sec-resolved}, we determine the explicit SUGRA solution corresponding to a heavy stack of
 D3-branes at a point on the resolved conifold, again by finding the Green's function on the manifold.
 We find a non-singular solution with an $AdS_5 \times S^5$ region and interpret this construction
 in gauge theory. We consider a wrapped Euclidean D3-brane to confirm the presence of baryonic VEVs and reproduce the wavefunction of a charged particle in a monopole field from the DBI action as a check on our calculations. We make a brief note on turning on a fluxless NS-NS $B_2$ field on the warped resolved
 conifold in Section \ref{sec-Bfield}. In Appendix A we discuss the harmonics on $T^{1,1}$ in
 co-ordinates that make
 the symmetries manifest. We then classify operators in the gauge theory by symmetry
 in an analogous way to enable simple matching of operator VEV's and normalizable fluctuations.

\section{The Conifold and its Resolution}
\label{sec-conifold}
\label{sec-smeared}

 The conifold is a singular non-compact Calabi-Yau three-fold \cite{Candelas}. Its importance
arises from the fact that the generic singularity in a Calabi-Yau
three-fold locally looks like the conifold. This is because it is
given by the quadratic equation,
 \beq
 z_1^2 + z_2^2 + z_3^2 + z_4^2 = 0.
 \eeq
 This homogeneous equation defines a real cone over a 5 dimensional manifold. For the cone to be
 Ricci-flat
the 5d base must be an Einstein manifold ($R_{\mu \nu} = 4 g_{\mu \nu}$).
 For the conifold \cite{Candelas}, the topology of the base can be shown to be $S^2 \times S^3$ and it is
 called $T^{1,1}$ with the following Einstein metric,
\begin{eqnarray} \label{T11metric}
  d\Omega_{T^{1,1}}^2 &=&  \frac{1}{9}  \left( d\psi + \cos \theta_1 d\phi_1 + \cos \theta_2 d\phi_2 \right)^2  \nonumber \\
 & &  \quad + \frac{1}{6}  (d\theta_1^2 + \sin^2 \theta_1 d\phi_1^2) + \frac{1}{6} (d\theta_2^2 + \sin^2 \theta_2 d\phi_2^2).
\end{eqnarray}
The metric on the cone is then $ds^2 = dr^2 + r^2
d\Omega_{T^{1,1}}^2$. As shown in \cite{Candelas} and earlier in
\cite{Romans}, $T^{1,1}$ is a homogeneous space, being the coset
$SU(2)\times SU(2)/U(1)$ and the above metric is the invariant
metric on the coset space.

We may introduce two other types  of complex coordinates on the
conifold, $w_a$ and $a_i,b_j$, as follows,
\begin{eqnarray} \label{coords}
 Z = \left( \begin{array}{cc}  z^3 + i z^4 & z^1 - i z^2 \\ z^1 +i z^2 & -z^3 +i z^4 \\ \end{array} \right) \nonumber
 = \left(\begin{array}{cc}w_1 & w_3 \\ w_4& w_2 \end{array} \right)
= \left( \begin{array}{cc}  a_1 b_1 & a_1 b_2 \\ a_2 b_1 & a_2 b_2 \\ \end{array} \right) \\
= r^\frac{3}{2} \left(\begin{array}{cc} - c_1 s_2 \; e^{\frac{i}{2} ( \psi + \phi_1-\phi_2)} &
    c_1 c_2 \; e^{\frac{i}{2} ( \psi+ \phi_1 +\phi_2)} \\
                  - s_1 s_2 \; e^{\frac{i}{2} ( \psi - \phi_1 -\phi_2)} & s_1 c_2 \;e^{\frac{i}{2}
                  ( \psi - \phi_1+\phi_2)} \end{array} \right)
\end{eqnarray}
where $c_i = \cos \frac{\theta_i}{2},\ s_i = \sin
\frac{\theta_i}{2}$ (see \cite{Candelas} for other details on the
$w,z$ and angular coordinates.) The equation defining the conifold
is now $\det Z = 0$.

The $a,b$ coordinates above will be of particular interest in this
paper because the symmetries of the conifold are most apparent in
this basis. The conifold equation has $SU(2)\times SU(2) \times
U(1)$ symmetry since under these symmetry transformations, \beq
\det L Z R^{T} = \det e^{i \alpha} Z = 0. \eeq This is also a
symmetry of the metric presented above where each $SU(2)$ acts on
$\theta_i,\phi_i,\psi$ (thought of as Euler angles on $S^3$) while
the $U(1)$ acts by shifting $\psi$. This symmetry can be
identified with $U(1)_R$, the R-symmetry of the dual gauge theory,
 in the conformal case.
The action of the $SU(2) \times SU(2) \times U(1)_R$ symmetry on
$a_i,b_j$ (defined in (\ref{coords})): \beqa SU(2) \times SU(2)
\mbox{  symmetry}&:& \quad
 \left( a_1 \atop a_2 \right)  \rightarrow L \; \left( a_1 \atop a_2 \right) ,
\qquad \left( b_1 \atop b_2 \right)  \rightarrow R \;
\left( b_1 \atop b_2 \right)  \\
 \mbox{R-symmetry}&:& \quad (a_i,b_j) \rightarrow e^{i \frac{\alpha}{2}} (a_i,b_j)
\ ,\eeqa i.e. $a$ and $b$ transform as $(1/2,0)$ and $(0,1/2)$
under $SU(2) \times SU(2)$ with R-charge $1/2$ each.
We can thus describe the singular conifold as points parametrized by
$a,b$ but from (\ref{coords}), we see that there is some redundancy
in the $a,b$ coordinates. Namely, the transformation \beq
\label{removeredun}
  a_i  \rightarrow \; \lambda \; a_i  \quad,\quad  b_j  \rightarrow \; \frac{1}{\lambda} \; b_j  \quad(\lambda \in \mathbf{C})
\eeq
 give the same $z,w$ in (\ref{coords}).
 We impose the constraint $|a_1|^2 + |a_2|^2 - |b_1|^2 -|b_2|^2 = 0$ to fix the magnitude in the
 above transformation. To account for the remaining phase, we describe the singular conifold as
 the quotient of the $a,b$ space with the above constraint by the relation
 $ a \sim e^{i\alpha} a, b \sim e^{-i\alpha} b$.

One simple way to describe the resolution is as the space obtained
by modifying the above constraint to, \beq \label{resconstr} |b_1|^2
+ |b_2|^2 - |a_1|^2 -|a_2|^2 = u^2 \eeq and then taking the
quotient, $ a \sim e^{i\alpha} a, b \sim e^{-i\alpha} b$. Then $u$
is a measure of the resolution and it can be seen that this space is
a smooth Calabi-Yau space where the singular point of the conifold
is replaced by a finite $S^2$. The complex metric on this space is
given in \cite{Candelas} while an explicit metric, first presented
in \cite{PandoZayas}, is: \beqa \label{resolvedmetric}
 ds_6^2 &=& \kappa^{-1}(r) dr^2 +  \frac{1}{9} \kappa(r) r^2 \left( d\psi + \cos \theta_1 d\phi_1 + \cos \theta_2 d\phi_2 \right)^2  \nonumber \\
 & &  \quad + \frac{1}{6} r^2 (d\theta_1^2 + \sin^2 \theta_1 d\phi_1^2) + \frac{1}{6} (r^2 + 6 u^2) (d\theta_2^2 + \sin^2 \theta_2 d\phi_2^2)
\eeqa
where
\beqa
  \kappa(r) = \displaystyle\frac{r^2 + 9 u^2}{r^2 + 6 u^2}\ ,
\eeqa 
where $r$ ranges from $0$ to $\infty$.
Note that the above metric has a finite $S^2$ of radius $u$ at
$r=0$, parametrized by $\theta_2,\phi_2$. Topologically, the resolved conifold
is an ${\bf R}^4$ bundle over $S^2$.
The metric asymptotes to that of the
singular conifold for large $r$.

Now we consider metrics produced by D3-branes on the conifold. As a
warm-up to the case of the resolved conifold, we consider the
example of placing a stack of D3-branes away from the apex of the
singular conifold. As in \cite{KW2}, the corresponding supergravity
solution is
\beqa \label{singmetric} ds^2
&=& \sqrt{H^{-1}(y)} \; \eta_{\mu\nu} dx^\mu dx^\nu + \sqrt{H(y)}
\left( dr^2 + r^2 d\Omega^2_{T^{1,1}} \right)\, ,\\ \label{f5dil}
F_5 &=& (1 +
*)dH^{-1}\wedge dx^0\wedge dx^1\wedge dx^2\wedge dx^3 , \qquad \Phi
= \mbox{const} \eeqa where $\mu, \nu = 0,1,2,3$ are the directions
along the D3-branes. $H(y)$ is a solution of the Green's equation on
the conifold
\beqa \label{greendef0} \Delta H(r,Z ; r_0, Z_0) =
\frac{1}{\sqrt{g}}\partial_m (\sqrt{g} g^{mn} \partial_n H ) &=& - {\cal C}
\frac{1}{\sqrt{g}} \delta(r-r_0) \delta^5(Z-Z_0) \ , \\
{\cal C} =2\kappa_{10}^2 T_3 N &=& (2 \pi)^4 g_s N (\alpha')^2 \ ,
\eeqa 
where
$(r_0,Z_0)$ is the location of the stack ($Z$ will represent
coordinates on $T^{1,1}$) and $T_3= \frac{1}{g_s (2\pi)^3 (\alpha')^2}$ is the
D3-brane tension.
%Here ${\cal C} =2\kappa_{10}^2 T_3 N=
% (2 \pi)^4 g_s N (\alpha')^2$
%(see \cite{BDKMMM}).

When the stack of D3-branes is placed at $r_0 = 0$, the solution is $H
=L^4/r^4$ where $L^4  = \frac{27 \pi g_s N (\alpha')^2}{4}$.  This reduces the
metric to ($z=L^2/r$),
\begin{eqnarray} \label{AdST11}
ds^2 = \frac{L^2}{z^2} \; (dz^2 + \eta_{\mu\nu} dx^\mu dx^\nu) +
L^2 d\Omega^2_{T^{1,1}}
\end{eqnarray}
This is the $AdS_5 \times T^{1,1}$ background, which is dual to the
superconformal $SU(N) \times SU(N)$ theory without any VEV's for the
bifundamental superfields. More general locations of the stack,
corresponding to giving VEV's that preserve the condition ${\cal
U}=0$, will be considered in section 4.

Now consider the case of resolved conifold. With D3-branes placed on
this manifold, we get the warped 10-d metric, \beq
\label{warpedresolved} ds_{10}^2 = \sqrt{H^{-1}(y)} dx^\mu dx_\mu +
\sqrt{H(y)} ds_6^2 \eeq where
$ds_6^2$ is the resolved conifold metric (\ref{resolvedmetric}) and
$H(y)$ is the warp factor as a function of the transverse
co-ordinates $y$, determined by the D3-brane positions. The dilaton
is again constant, and $F_5 = (1 +
*)dH^{-1}\wedge dx^0\wedge dx^1\wedge dx^2\wedge dx^3$.

In \cite{PandoZayas}, the warped supergravity solution was worked
out assuming a warp factor with only radial dependence (i.e no
angular dependence on $\theta_2,\phi_2$):
\begin{equation} H_{PT}(r)= \frac{2L^4}{9u^2 r^2} - 
\frac{2L^4}{81 u^4} \log \left(1 + \frac{9u^2}{r^2}\right )\ .
\end{equation}
 The small $r$ behavior of
$H_{PT}$ is $\sim \frac{1}{r^2}$.
This produces a metric singular at $r=0$ since the radius of
$S^2(\theta_2,\phi_2)$ blows up and the Ricci tensor is singular.
 Imposing the symmetry that $H$ has only radial dependence corresponds not to having a stack of
 D3-branes at a point (which would necessarily break the $SU(2)$ symmetry in $\theta_2,\phi_2$)
 but rather having the branes smeared out uniformly on the entire two sphere at the origin.
 The origin of this singularity is precisely the smearing of the D3-brane charge.
 In Section~\ref{sec-resolved}, we confirm this by constructing the solution corresponding to
 localized branes and find that there is no singularity.

\section{Flows on the Singular Conifold}
 \label{sec-singular}

Let us consider the case when the stack of D3-branes is moved away
from the singular point of the conifold. Since the branes are at a
smooth point on the conifold, we expect the near brane geometry to
become $AdS_5 \times S^5$ and thus describe ${\cal N} = 4$ $SU(N)$
SYM theory. The warp factor $H(r,Z)$ can be written as an expansion
in harmonics on $T^{1,1}$ starting with the leading term $1/r^4$
followed by higher powers of $1/r$. Thus, the full solution still
looks like $AdS_5 \times T^{1,1}$ at large $r$, but further terms in
the expansion of the warp factor change the geometry near the branes
to $AdS_5 \times S^5$. Such a SUGRA solution describes the RG flow
from the ${\cal N} = 1$ $ SU(N) \times SU(N)$ theory in the UV to
the ${\cal N}=4$ $SU(N)$ SYM in the IR. We will confirm this
explicitly through the computation of the general Green's function
on the conifold. We display the series of perturbations of the
metric and interpret these normalizable solutions in terms of
 VEVs in the gauge theory for a series of operators using the setup of Appendix A.
This was originally studied in \cite{KW2} where a restricted class of chiral
operators was considered.

\begin{figure}[t]
\centering
\includegraphics{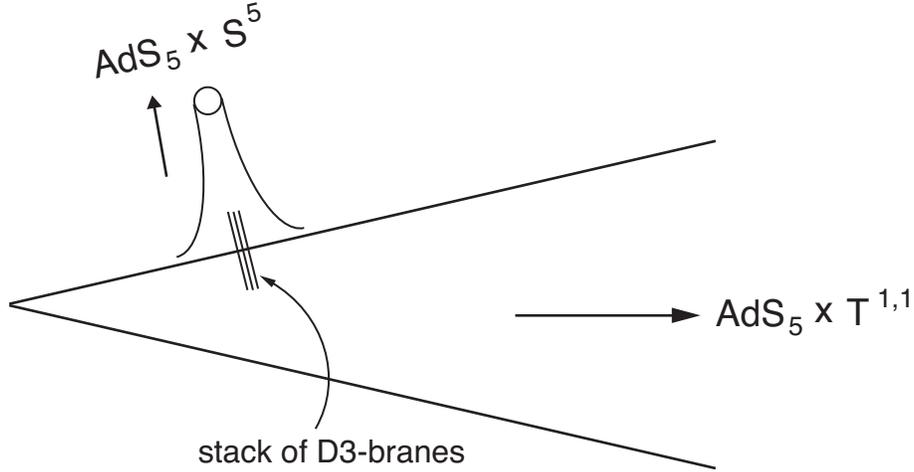}
\caption{A stack of D3-branes warping the singular conifold}
\end{figure}

Let us place the stack at a point $(r_0,Z_0)$ on the singular
conifold. We rewrite (\ref{greendef0}) as
\begin{eqnarray}
\label{Green}
\label{SGf}
\Delta H = {\Delta_r} H + \frac{ \Delta_Z}{r^2} H & = & - \frac{{\cal C}}{\sqrt{g}}\delta( r - r_0)\Pi_i \delta^5(Z_i -Z_{0i}) \nonumber \\
& \equiv& - \frac{{\cal C}}{\sqrt{g_r}} \delta( r - r_0) \delta_{A}(Z - Z_0)
\end{eqnarray}
where $\Delta_r = \frac{1}{\sqrt{g}} \: \partial_r \left(\sqrt{g} \: \partial_r \right)$ is the radial Laplacian, $\Delta_{Z}$ the remaining angular laplacian. In the second line, $g_r$ is defined to have the radial dependence in $g$ and the angular delta function $\delta_{A}(Z - Z_0)$ is defined by absorbing the angular factor $\sqrt{g_5} = \sqrt{g/g_r}$. In this section, we have $ \sqrt{g} = \frac{1}{108} r^5 \sin \theta_1 \sin \theta_2$ and we take $\sqrt{g_r} = r^5$.

The eigenfunctions $Y_I(Z)$ of the angular laplacian on $T^{1,1}$ can be classified by a set $I$
of symmetry charges since $T^{1,1}$ is a coset space \cite{Gubser,Ceresole}. The eigenfunctions $Y_I$ are constructed explicitly in the appendix, including using the $a_i,b_j$ coordinates which will facilitate the comparison with the gauge theory below.
If we normalize these angular eigenfunctions as,
\begin{eqnarray}
\label{Ynorm}
\int \; Y^*_{I_0}(Z) Y_I(Z) \;\; \sqrt{g_5} \; d^5 \varphi_i & = & \delta_{I_0, I}
\end{eqnarray}
we then have the complementary result,
\begin{eqnarray}
\label{deltaexp}
\sum_I Y^*_I(Z_0) Y_I(Z) = \frac{1}{\sqrt{g_5}} \; \delta(\varphi_i - \varphi_{0i}) \equiv \delta_{A}(Z - Z_0).
\end{eqnarray}

We expand the $\delta_{A}(Z - Z_0)$ in (\ref{Green}) using (\ref{deltaexp}) and see that the Green's function can be expanded as,
\begin{eqnarray}
\label{Greenexp}
  H & = & \displaystyle\sum_I  \; H_I(r, r_0) \; Y_I(Z) \; Y^*_I(Z_0)
\end{eqnarray}
which reduces (\ref{Green}) to the radial equation,
\begin{eqnarray}
\displaystyle{ \label{eq2}
 \frac{1}{r^5} \frac{\partial}{\partial r} \left( r^5 \frac{\partial}{\partial
r} H_I\right) -  \frac{E_I}{r^2} H_I = - \frac{{\cal C}}{r^5}\delta( r - r_0)
}
\end{eqnarray}
where $ \Delta_Z Y_I(Z) = - E_I Y_I(Z)$ (see appendix A for details of $E_I$.)

As is easily seen, the solutions to this equation away from $ r = r_0$ are

$$
H_I \; = \; {A_\pm \; r^{c_\pm}}, \textrm{ where } \; \; c_\pm = -2\pm \sqrt{E_I +4}.
 $$
The constants $A_\pm$ are uniquely determined  integrating (\ref{eq2}) past
$r_0$. This determines $H_I$ and we put it all together to get the
solution to (\ref{SGf}), the Green's function on the singular
conifold
\begin{equation} \label{singgreen}
H(r ,Z ; r_0, Z_0) =   \sum_I \frac{{\cal C}}{2 \sqrt{E_I +4} } \;
Y^*_I(Z_0) Y_I(Z) \; \times
 \left\{ \begin{array}{ll}
  \displaystyle{\frac{1}{r^4_0} \left(\frac{r}{r_0}\right)^{c_I} }& r \leq r_0 \\
 \\
  \displaystyle{\frac{1}{r^4} \left(\frac{r_0}{r}\right)^{c_I} } & r \geq r_0 \ ,\\
\end{array} \right.
\end{equation}
where $c_I = c_+$. The term with $E_I=0$ gives $L^4/r^4$ where 
\begin{equation}
L^4 = \frac{{\cal C} }{4 \mathbf{Vol}(T^{1,1})}=
\frac{27 \pi g_s N (\alpha')^2}{4}
\ .\end{equation}
Since $E_I = 6(l_1(l_1+1) + l_2(l_2+1) - R^2/8)$, there are $(2l_1+1) \times (2l_2+1)$ terms
with the same $E_I$ and hence powers of $r$ and factors.
Also note that when $l_1 = l_2 = \pm \frac{R}{2}$, $c_I$ is a rational number and
these are related to (anti) chiral superfields in the gauge theory.

We can argue that the geometry near the stack (at $r_0,Z_0$) is actually a long 
$AdS_5 \times S^5$ throat.
We observe that $H$ must behave as $L^4/y^4$ near the stack 
(where $y$ is the distance between $(r,Z)$ and $(r_0,Z_0)$) since it is the solution 
of the Green's function and locally, the manifold looks flat and is 6 dimensional.
This leads to the usual $AdS_5\times S^5$ throat. We show this explicitly in Appendix B.
The complete metric thus describes holographic RG flow from
  $AdS_5 \times T^{1,1}$  geometry in the UV to $AdS_5 \times S^5$ in the IR. 
Note, however that
this background has a conifold singularity at $r=0$.

\subsection*{Gauge theory operators}
Let the stack of branes be placed at a point $a_i,b_j$ on the conifold. Then consider assigning the VEVS, $A_i = a_i^* 1_{N\times N}, B_j=b_j^* 1_{N\times N}$, i.e the prescription
 \beq \label{singvevs}
 Z_0  = \left( \begin{array}{cc}  a_1 b_1 & a_1 b_2 \\ a_2 b_1 & a_2 b_2 \\ \end{array} \right)  \quad \Longleftrightarrow  \quad \begin{array}{cc}
A_1 = a_1^* 1_{N\times N}, & A_2 = a_2^* 1_{N\times N}, \\ B_1 = b_1^* 1_{N\times N}, & B_2 = b_2^* 1_{N\times N}.  \end{array}
 \eeq
 In the appendix, we construct operators ${\cal O}_I$ transforming with the symmetry charges $I$. From the similar construction of the operator ${\cal O}_I$ and $Y_I(Z)$ (compare (\ref{compwvf}) and  (\ref{operatorL})), this automatically leads to a VEV proportional to $Y^*_I(Z_0)$ for the operator ${\cal O}_I$.

Meanwhile, the linearized perturbations
of the metric are determined by binomially expanding  $\sqrt{H}$ in (\ref{singmetric})
and considering terms linear in $Y_I(Z)$. These are easily seen to be of the form $
Y^*_I(Z_0) Y_I(Z) { \left(\frac{r_0}{r}\right)^{c_I} }$.
 From its form and symmetry properties, we conclude that it is the dual to the above VEV,
\beqa
 Y^*_I(Z_0) Y_I(Z) \; \times  \displaystyle{ \left(\frac{r_0}{r}\right)^{c_I} }  \quad \Longleftrightarrow \quad  \langle {\cal O}_I \rangle \propto Y^*_I(Z_0) \; r_0^{c_I}.
\eeqa

This is the sought relation between normalizable perturbations and operator VEV's.
For a general position of the stack $(r_0,Z_0)$, all $Y_I^*(Z_0)$ are non-vanishing. Being a coset space, we can use the symmetry of $T^{1,1}$, to set the D3-branes to lie at any specific point without loss of generality.
For example, consider the choice
\beq
Z_0 = \left( \begin{array}{cc}  a_1 b_1 & a_1 b_2 \\ a_2 b_1 & a_2 b_2 \\ \end{array} \right)  = \left( \begin{array}{cc} 1 & 0 \\ 0 & 0 \end{array} \right) \qquad \Rightarrow \qquad a_1 = b_1 = 1 , \; a_2 = b_2 = 0.
\eeq

Using (\ref{compwvf}) and (\ref{wvfdefn}) for $Y_I$, we find that $Y_I(Z_0) = 0$ unless $m_1 = m_2 = R/2$ and for these non-vanishing $Y_I$ we get,
\beq \label{specialfluct}
Y_I(Z_0) \sim  a_1^{l_1+\frac{R}{2}} \bar{a}_1^{l_1-\frac{R}{2}} b_1^{l_2+\frac{R}{2}} \bar{b}_1^{l_2-\frac{R}{2}}
\eeq
 If we give the VEVs $A_1 = B_1 = 1_{N\times N}, A_2 = B_2 = 0$, 
we get $\langle TrA_1 B_1\rangle \neq 0$ and all other $\langle TrA_i B_j \rangle = 0$. 
In fact, by this assignment, the only gauge invariant operators with non-zero vevs are 
the ${\cal O}_I$ with $m_1=m_2 = R/2$. These are precisely the operators dual to 
fluctuations $Y_I(Z)$ that have non-zero coefficient $Y_I^*(Z_0)$ as was seen in 
(\ref{specialfluct}).

 The physical dimension of this operator (at the UV fixed point) is read off as $c_I$ from the metric fluctuation - a supergravity prediction for strongly coupled gauge theory. (Above, $r_0$ serves as a scale for dimensional consistency.) In \cite{KW2}, the (anti) chiral operators were discussed ($l_1 = l_2 = \pm \frac{R}{2}$) . These have rational dimensions but as we see here, for any position of the stack of D3-branes, other operators (with generically irrational dimensions) also get vevs. For example, the dimension of the simplest non-chiral operator ($I \equiv l_1 =1, l_2 =0, R =0$) is $2$ but when $I \equiv l_1=2,l_2=0, R=0$, ${\cal O}_I$ has dimension $2(\sqrt{10}-1)$. This interesting observation about highly non-trivial scaling dimensions in strongly coupled gauge theory was first made in \cite{Gubser}.

 When operators $A_i,B_j$ get vevs as in (\ref{singvevs}), the $SU(N) \times SU(N)$ gauge
group is broken down to the diagonal $SU(N)$. The bifundamental fields
$A,B$ now become adjoint fields. With one linear combination of fields having a VEV,
we can expand the superpotential
$W \sim {\rm Tr} \det A_i B_j =
{\rm Tr} (A_1 B_1 A_2 B_2 - A_1 B_2 A_2 B_1)$ of the $SU(N)\times SU(N)$ theory to find that it
is of the form ${\rm Tr} (X[Y,Z])$ in the remaining adjoint fields \cite{KW}.
This is exactly ${\cal N} = 4$ $SU(N)$ super Yang-Mills, now obtained through symmetry breaking in
the conifold theory. This corresponds to the $AdS_5 \times S^5$ throat we found on the
gravity side near the source at $r_0,Z_0$.

 Thus we have established a gauge theory RG flow from ${\cal N}=1$ $SU(N) \times SU(N)$ theory in the UV to
${\cal N}=4$ $SU(N)$ theory in the IR. The corresponding gravity dual was constructed and found to be asymptotically $AdS_5 \times T^{1,1}$ (the UV fixed point) but developing a $AdS_5 \times S^5$ throat at the other end of the geometry (the IR fixed point). The simple example is generalized to the resolved conifold in the next section.

\section{Flows on the Resolved Conifold}
\label{sec-resolved}
In this section we use similar methods to construct the Green's function on the resolved conifold and corresponding
warped solutions due to a localized stack of D3-branes.
We will work out explicitly
the $SU(2)\times U(1)\times U(1)$ symmetric RG flow
corresponding to a stack of D3-branes localized on the finite $S^2$ at $r=0$.
Such a solution is dual to giving a VEV to just one bi-fundamental field, e.g. $B_2$, which
Higgses the ${\cal N} = 1$ $SU(N)\times SU(N)$ gauge theory theory to
the ${\cal N}=4$ $SU(N)$ SYM. We also show
how the naked singularity found in \cite{PandoZayas} is removed through the localization of the D3-branes.

  The supergravity metric is of the form (\ref{warpedresolved}).
The stack could be placed at non-zero $r$ but in this case,
the symmetry breaking
pattern is similar in character to the singular case discussed above.
The essence of what is new to the resolved conifold is best captured with the stack placed at
a point on the
blown up $S^2$ at $r=0$; this
breaks the $SU(2)$ symmetry rotating $(\theta_2,\phi_2)$ down to a $U(1)$.
The branes also preserve the $SU(2)$ symmetry rotating $(\theta_1,\phi_1)$ as well as the
$U(1)$ symmetry corresponding to the shift of $\psi$.
On the other hand, the $U(1)_B$ symmetry is broken because
the resolved conifold has no non-trivial three-cycles \cite{KW2}. Thus the warped
resolved conifold background has
unbroken $SU(2) \times U(1) \times U(1)$ symmetry.

To match this with the gauge theory, we first recall that in the absence of VEV's
we have $SU(2) \times SU(2) \times U(1)_R \times U(1)_B$ where the $SU(2)$'s act on
$A_i,B_j$ respectively, the $U(1)_R$ is the R-charge ($R_A = R_B = 1/2$) and
$U(1)_B$ is the baryonic symmetry, $A \to e^{i \theta} A, B \to e^{-i \theta} B$.
As noted above, the VEV $B_2 = u 1_{N\times N}, B_1 = A_i = 0$ corresponds to placing the branes
at a point on the blown-up 2-sphere. This clearly leaves one of the $SU(2)$ factors unbroken.
While $U(1)_R$ and $U(1)_B$ are both broken by the baryonic operator $\det B_2$,
their
certain $U(1)$ linear combination remains unbroken.
Similarly, a combination of $U(1)_B$ and the
$U(1)$ subgroup of the other $SU(2)$, that rotates the $B_i$ by phases,
remains unbroken.
Thus we again have $SU(2) \times U(1) \times U(1)$ as the unbroken symmetry,
consistent with the warped resolved conifold solution.
  Since the baryon operator $\det B_2$ acquires a VEV while no chiral
mesonic operators do (because $A_1=A_2=0$), the solutions found in this section are dual to
a ``baryonic branch'' of the CFT (see \cite{Benvenuti} for a discussion of such branches).

\subsection*{Solving for the warp factor}

Since the resolution of the conifold preserves the $SU(2)_L\times SU(2)_R\times U(1)_\psi$ symmetry (where $U(1)_\psi$ shifts $\psi$),
the equation for Green's function $H$ looks analogous to (\ref{SGf}) for the resolved conifold,
\begin{equation}
\label{RGf}
\frac{1}{r^3(r^2+6 u^2)} \frac{\partial}{\partial r} \left( r^3(r^2+6 u^2) \kappa(r) \frac{\partial}{\partial r}
H\right) +  \mathbf{A} H   =
 - \frac{{\cal C}}{r^3(r^2 + 6 u^2)}\delta( r - r_0) \;\; \delta_{T^{1,1}}(Z - Z_0)
\end{equation}
where
\begin{equation}
\label{reseq2}
\mathbf{A} H  =  6 \frac{\Delta_1}{r^2} H  +  6 \frac{\Delta_2}{r^2 + 6 u^2} H+ 9 \frac{\Delta_R}{\kappa(r) r^2}  H
\end{equation}
and $\Delta_i$ , $\Delta_R$ are defined in the appendix. ($\Delta_i$ are $S^3$ laplacians and
$\Delta_R = \partial_\psi^2$. Note that $6 \Delta_1 + 6 \Delta_2 + 9 \Delta_R = \Delta_{T^{1,1}}$).
%Again, the normalization factor $C= (2 \pi)^4 g_s N (\alpha')^2$.

 This form of the ${\bf A}$ is fortuitous and allows us to use the $Y_I$ from the singular conifold,
since $Y_I$ are eigenfunctions of each of the three pieces of ${\bf A} $ above.
We could solve it for general $r_0$, but $r_0 = 0$ is a particularly simple case that is of primary interest
in this paper.

Since (\ref{RGf}) involves the same $ \delta_{T^{1,1}}(Z - Z_0)$ as the singular case,
we can expand $H$ again in terms of the angular harmonics and radial functions as $H = \sum_{I}  H_I(r,r_0)
Y_I(Z) Y^*_I(Z_0)$ to find the radial equation,
\begin{eqnarray}
 \label{resradialeqn}
&-& \frac{1}{r^3(r^2+6 u^2)} \frac{\partial}{\partial r} 
\left( r^3 (r^2 + 6 u^2) \kappa(r) \frac{\partial}{\partial
r} H_I\right)  \nonumber \\
\quad &+& \left( \frac{6 (l_1(l_1+1) - R^2/4)}{r^2} +
\frac{6 (l_2(l_2+1) - R^2/4)}{r^2 + 6 u^2} + \frac{9 R^2/4}{\kappa(r)r^2} \right) H_I =  
\frac{{\cal C}}{r^3(r^2+6 u^2)}\delta(r - r_0) .
\end{eqnarray}
This equation can be solved for $H_I(r)$ exactly in terms of some special functions. If we place the
stack at $r_0 = 0$, i.e at location $(\theta_0,\phi_0)$ on the blown up $S^2$, then an additional
simplification occurs.
The warp factor $H$ must be a singlet under the $SU(2)\times U(1)_\psi$ that rotates $(\theta_1,\phi_1)$ and $\psi$ since
these have shrunk at the point where the branes are placed.
Hence we only need to solve this equation for $l_1= R = 0$, $l_2=l$.

The two independent solutions (with convenient normalization) to the homogeneous 
equation in this case,
in terms of the hypergeometric function ${}_2F_1$, are
\beqa \label{resradialsoln}
H_l^A(r) & = & \frac{2 }{9u^2} \frac{C_\beta}{r^{2+2\beta}} \;\; 
{}_2F_1\left(\beta,1+\beta ;1+ 2
\beta;-\frac{9 u^2}{r^2}\right)  \nonumber \\
H_l^B(r) &\sim & {}_2F_1 \left(1-\beta,1+\beta;2;-\frac{r^2}{9u^2}\right)
\eeqa
where
\beqa \label{normfactor}
C_\beta = \frac{(3u)^{2\beta} \Gamma(1+\beta)^2}{\Gamma(1+2 \beta)} \ ,\qquad 
\beta = \sqrt{1+(3/2)l(l+1)}\ .
\eeqa
These two solutions have the following asymptotic behaviors,
\beqa
\frac{2}{9 u^2 r^2}+ \frac{4 \beta^2}{81 u^4} \ln r + {\cal O}(1)
 \quad \stackrel{0\leftarrow r}{\longleftarrow} 
\quad &H_l^A(r)& \quad \stackrel{r
\rightarrow \infty}{\longrightarrow} \quad \frac{2  C_\beta}{9 u^2 r^{2 + 2\beta}} \\
\nonumber \\
{\cal O}(1) \quad \stackrel{0\leftarrow r}{\longleftarrow} \quad &H_l^B(r) &\quad
\stackrel{r \rightarrow \infty}{\longrightarrow} \quad 
{\cal O}\left (r^{- 2 + 2\beta}\right )
\eeqa

To find the solution to (\ref{resradialeqn}) 
with the $\delta(r-r_0)$ on the RHS, we need to match the 
two solutions at $r=r_0$ as well as satisfy the condition on 
derivatives obtained by integrating past $r_0$. Since we are interested 
in normalizable modes, we use $H_l^A(r)$ for $r > r_0$ and $H_l^B(r)$ for $r<r_0$.
Finally, we take $r_0 = \epsilon$ and take the limit $\epsilon \rightarrow 0$ 
(since the stack of branes is on
the finite $S^2$). We find simply that $H_l(r) = {\cal C} H_l^A(r)$ due to the 
normalization chosen earlier in (\ref{resradialsoln}).
Putting it all together, we find,
\beqa \label{resfullsoln}
   H(r,Z;r_0=0,Z_0)  & = & {\cal C} \sum_I Y^{*}_I(Z_0) H_I^A(r) Y_I(Z)
\eeqa
where only the $l_1=0, R=0$ harmonics
contribute since the stack leaves $SU(2) \times U(1)\times U(1)$ symmetry unbroken.
In this situation, the $Y_I$ wavefunctions simplify to the usual $S^2$ spherical 
harmonics $\sqrt\frac{4 \pi}{\mathbf{Vol}(T^{1,1})} \; Y_{l,m}$.

 Let us take $b_i$ to describe the finite $S^2(\theta_2,\phi_2)$ while $a_j$ are associated with the $S^2$
that shrinks to a point. As reviewed in Section \ref{sec-conifold},
the resolved conifold can be described with $a,b$ variables governed by the constraint (\ref{resconstr}),
where $u$ is the measure of resolution, the radius of the finite $S^2$.
The position of the branes on the finite sphere can be parametrized as
$b_1 = u \sin \frac{\theta_0}{2} e^{-i \phi_0/2},
b_2 = u \cos \frac{\theta_0}{2} e^{i \phi_0/2}$ and $a_1=a_2=0$
(since the branes do not break the $SU(2)$ symmetry rotating
the $a$'s).
Then,
\beqa \label{resolvedgreen}
    H(r,Z;r_0=0,Z_0=(\theta_0,\phi_0))  = 
4 \pi L^4 \sum_{l, m} H_l^A(r) \; Y^*_{l,m}(\theta_0, \phi_0)
 Y_{l,m}(\theta_2, \phi_2).
\eeqa
Without a loss of generality, we can place
the stack of D3-branes at the north pole ($\theta_0 = 0$) of the 2-sphere. 
Then (\ref{resolvedgreen}) simplifies further: 
only $m=0$ harmonics contribute and we get the explicit 
expression for the warp factor which is one of our main results,
\beqa \label{simpwarp}
    H(r,\theta_2)  =   
L^4 \sum_{l=0}^\infty  \; (2 l + 1) H_l^A(r) P_l(\cos \theta_2).
\eeqa
Now the two unbroken $U(1)$ symmetries are manifest as shifts of $\phi_2$ and $\psi$.

The `smeared' singular solution found in \cite{PandoZayas}
corresponds to retaining only the $l=0$ term in this sum. Indeed, we
find that \beq \label{zmode}
H_0^A(r) =  \frac{2 C_1}{9 u^2 r^{4}} \;\;
{}_2F_1\left(1,2 ;3 ;-\frac{9 u^2}{r^2}\right) = \frac{2}{9u^2 r^2}
- \frac{2}{81 u^4} \log \left(1 + \frac{9u^2}{r^2}\right) \eeq in
agreement with \cite{PandoZayas}. Fortunately,
if we consider the full sum over modes appearing in
(\ref{resolvedgreenN}), the geometry is no longer singular. 
The leading term in the warp factor (\ref{simpwarp}) at small $r$ is
\beqa \label{resolvedgreenN}
{2L^4\over 9u^2 r^2}  
\sum_{l=0}^\infty  \; (2 l + 1) P_l(\cos \theta_2)= {4 L^4\over 9 u^2 r^2}
\delta ( 1-\cos \theta_2) 
\eeqa
This shows that away from the north pole the $1/r^2$ divergence of the warp factor
cancels. Similarly, after summing over $l$
the term $\sim \ln r$ cancels away from the north pole. This implies that
the warp factor is finite at $r=0$ away from the north pole.
However, at the north pole it diverges as expected.
 Indeed, since the branes are now localized at a smooth point on 
the 6-manifold (all points on the resolved conifold are smooth), 
very near the source $H$ must again be of the form $L^4/y^4$ where 
$y$ is the distance from the source. This is shown explicitly in Appendix B.
Writing the local metric in the form $dy^2 + y^2 d\Omega^2_{S^5}$ near the source, we
get the $AdS_5 \times S^5$ throat, avoiding the singularity found in \cite{PandoZayas}.

\begin{figure}[t]
\centering
\includegraphics{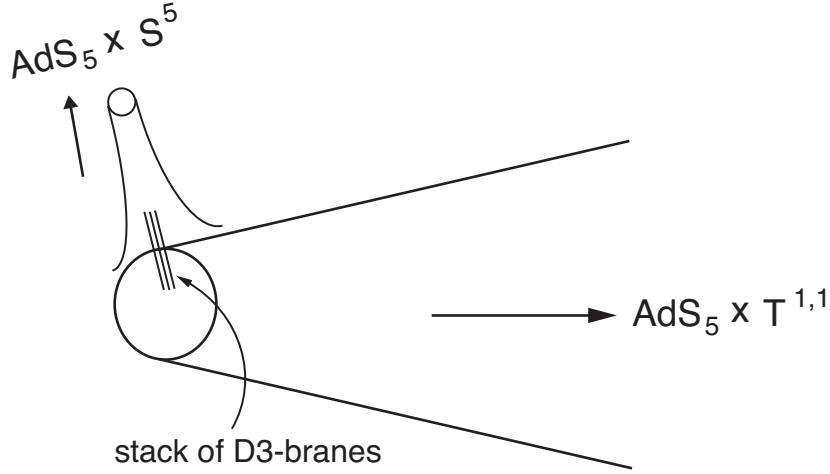}
\caption{A stack of D3-branes warping the resolved conifold}
\end{figure}

\subsection*{ Gauge theory operators}
With the branes placed at the point $b_1 = u \sin \frac{\theta_0}{2} e^{-i \frac{\phi_0}{2}}, 
b_2 = u \cos \frac{\theta_0}{2} e^{i \frac{\phi_0}{2}}$ , $a_1=a_2=0$ on the finite $S^2$, 
consider the assignment of VEVs, $B_1 =  u \sin \frac{\theta_0}{2} e^{i \frac{\phi_0}{2}} 1_{N\times N} , 
B_2 = u \cos \frac{\theta_0}{2} e^{-i \frac{\phi_0}{2}} 1_{N\times N}, A_1 = A_2 = 0$.

The linearized fluctuations compared to the leading term $1/r^4$ ($l_2 = 0$) are of the form,
\beqa \label{resfluctdim}
Y_I^*(Z_0) Y_I(Z) r^4 H_I(r)  \rightarrow  Y_I^*(Z_0) Y_I(Z) (\frac{1}{r})^{c_I} \quad ( r \gg a)
\eeqa
where as earlier, $c_I = 2 \sqrt{1+(3/2)l_2(l_2+1)} - 2$.

With this assignment of VEVs above, operators ${\cal O}_I$ with $l_1
= R = 0$ acquire VEVs. In the notation $|l_2,m_2;R>$ of Appendix A,
\beqa \label{resoperatorvevs} <{\cal O}_I> &=& <Tr
|l_2,m_2;0\rangle_B  > \neq 0. \eeqa For example, when $l_2 =
1,m_2=0$ above, $\langle{\cal O}_I\rangle= \langle Tr B_1 \bar{B}_1
- B_2 \bar{B}_2\rangle = u^2( \sin^2 \frac{\theta_0}{2} - \cos^2
\frac{\theta_0}{2}) $. By construction of ${\cal O}_I$ and $Y_I$, it
is clear that $<{\cal O}_I> \sim Y_I^*(Z_0)$ and is dual to the
metric fluctuation above. We can read off the dimensions of these
operators as $c_I$
% = 2\beta -2, \beta = \sqrt{1+ (3/2)l_2(l_2+1)}$
from the large $r$ behavior of the fluctuation (\ref{resfluctdim}).
For the $l_2=1$ operator as above, the exact dimension is $2$ (the
classical value) because the operator is a superpartner of a conserved
current \cite{Itzhaki}. Similarly, the dimension 2 of ${\cal U}$
is protected against quantum corrections 
because of its relation to a conserved baryonic current \cite{KW2}.
  When one expands $H_I(r)$ at large $r$, one finds sub-leading terms in addition to
$1/r^{c_I}$ shown above. These terms, which do not appear for the singular conifold,
increase in powers of $1/r^2$ and hence describe a series of operators with
the same symmetry $I$ but dimension increasing in steps of 2 from $c_I$.
These modes appear to correspond to VEV's for the operators
$ {\rm Tr}  {\cal O}_I {\cal U}^n $. It would be interesting to 
investigate such operators
and their dimensions further.

Hence we have an infinite series of operators that get VEV's in
the gauge theory dual to the warped resolved conifold.
These are in addition to the basic operator ${\cal U}$ which gets
a VEV
due to the asymptotics of the unwarped resolved conifold
metric itself \cite{KW2}. 
The operator ${\cal U}$ would get the same VEV of 
$u^2$ for any position of the brane on the $S^2$ while the VEV's
for the infinite series of operators ${\cal O}_I$ depend on the position.
We also note that ${\cal U} = u^2$ is the gauge dual of the
constraint $|b_1|^2 + |b_2|^2 - |a_1|^2 - |a_2|^2 = u^2$ 
defining the resolved conifold in Section
\ref{sec-conifold}.

Lastly, we verify that the gauge theory does flow in the infrared to ${\cal N}=4$ $SU(N)$ SYM.
Without loss of generality, we can take the stack of branes to lie on the north pole of the finite sphere
($B_2 = u 1_{N \times N}, B_1 = 0$). As in the singular case, $B_2 = {u} 1_{N\times N}$ breaks the
$SU(N) \times SU(N)$ gauge group down to $SU(N)$, all the chiral fields now transforming in the
adjoint of this diagonal group. Consider the ${\cal N} =1$ superpotential
$W \sim {\rm Tr} \det A_i B_j = Tr (A_1 B_1 A_2 B_2 - A_1 B_2 A_2 B_1)$.
When $B_2 \propto 1_{N\times N}$,
the superpotential reduces to the $ {\cal N}=4$ form,
  \beq
  W = \lambda  {\rm Tr} (A_1 B_1 A_2 - A_1A_2 B_1) = \lambda {\rm Tr} (A_1[B_1,A_2]).
  \eeq
This confirms that the gauge theory flows to the $ {\cal N}=4$ $SU(N)$ SYM theory in the infrared.

\subsection*{Baryonic Condensates and Euclidean D3-branes}

Here we present a calculation of the baryonic VEV using the dual string theory on
the warped resolved conifold background.\footnote{We are indebted to E. Witten
for his suggestion that led to the calculation presented in
this section.} 
A similar question was addressed for the
cascading theories on the baryonic branch where the baryonic condensates are
related to the action of a Euclidean D5-brane wrapping
the deformed conifold \cite{Wittenunpub,BDK}. 
In this section we present an analogous construction for the
warped resolved conifolds, which are asymptotic to $AdS_5\times T^{1,1}$. 

The objects in $AdS_5\times T^{1,1}$ that are dual to baryonic operators
are D3-branes wrapping 3-cycles in $T^{1,1}$ \cite{GK}. Classically, the
3-cycles dual to the baryons made out of the $B$'s are located at fixed
$\theta_2$ and $\phi_2$ (quantum mechanically, one has to carry out
collective coordinate quantization and finds wave functions of spin $N/2$
on the 2-sphere). To calculate VEV's of such baryonic operators, we need
to consider Euclidean D3-branes which at large $r$ wrap a 3-cycle at fixed $\theta_2$
and $\phi_2$. In fact, the symmetries of the calculation suggest that the
smooth 4-cycle wrapped by the Euclidean D3-brane is located at fixed
$\theta_2$ and $\phi_2$, and spans the $r$, $\theta_1$, $\phi_1$ and $\psi$ directions. 
In other words, the Euclidean D3-brane wraps the ${\bf R}^4$ fiber of the 
${\bf R}^4$ bundle over $S^2$ (recall that the resolved conifold is such a bundle).

The action of the D3-brane will be integrated up to a radial cut-off $r_c$, and
we identify $e^{-S(r_c)}$ with the classical field dual to the baryonic operator.
The Born-Infeld action is
\beq
S_{BI}= T_3 \int d^4\xi \sqrt{ g} 
\ ,
\eeq
where $g_{\mu\nu}$ is the metric induced on the D3 world volume. 
%Reinstating the normalization factor $L^4 = \frac{27 \pi g_s N (\alpha')^2}{4}$,
We find 
\beq
S_{BI}= 
{3N\over 4L^4}\int_0^{r_c} dr r^3 H(r,\theta_2)= 
{3N\over 4}\int_0^{r_c} dr r^3  
\sum_{l=0}^\infty  \; (2 l + 1) H_l^A(r) P_l(\cos \theta_2)
\ .
\eeq
%where $H(r,\theta_2)$ is the warp factor normalized as in 
%(\ref{simpwarp}). 

The $l=0$ term (\ref{zmode}) needs to be evaluated separately since it contains
a logarithmic divergence:\footnote{A careful holographic renormalization of
divergences for D-brane actions was considered in \cite{Karch}. 
We leave a similar construction in the present situation for
future work.}
\beq
\int_0^{r_c} dr r^3 H_0^A (r)= {1\over 4} + {1\over 2}
\ln \left (1+ {r_c^2\over 9 u^2} \right )
\ .
\eeq
For the $l>0$ terms the cut-off may be removed and we find
a nice cancellation involving the normalization (\ref{normfactor}):
\beq
\int_0^{\infty} dr r^3 H_l^A (r)= {2\over 3 l(l+1)}\ .
\eeq
Therefore,
\beq
\int_0^{\infty} dr r^3 \sum_{l=1}^\infty (2l+1) H_l^A (r) P_l (\cos \theta_2)
= {2\over 3}\sum_{l=1}^\infty  {2l+1\over l(l+1)} P_l (\cos \theta_2)
={2\over 3} (-1 -2 \ln [\sin (\theta_2/2)])\ .
\eeq
This expression is recognized as the Green's function on a sphere.
Combining the results, and taking $r_c\gg u$, we find
\beq
e^{-S_{BI}} = \left ( {3 e^{5/12} u\over r_c }\right )^{3N/4} 
\sin^N (\theta_2/2)\ .
\eeq
%This is not the complete answer since we need to 
%include the Chern-Simons term describing
%the coupling of the D3-brane to the background $C_4$. Studying the structure of the
%self-dual 5-form field
%\beq
%F_5 = (1+ *) d H^{-1}\wedge dx^0\wedge dx^1\wedge dx^2\wedge dx^3
%\eeq 
%we find thatg there exists a component of 
%$C_4$ multiplying $d\theta_1\wedge d\phi_1\wedge d\psi\wedge dr$.
%While the integral of $C_4$ could be evaluated directly, we may use the
%holomorphy arguments similar to those in \cite{BDKMMM} to argue that
%$e^{-S_{CS}}= e^{-i N\phi_2/2}$, so that
%\beq
%e^{-S_{BI}-S_{CS}} = \left ( {3 e^{5/12} u\over r_c }\right )^{3N/4} 
%\sin^N (\theta_2/2) e^{-iN\phi_2/2}\ .
%\eeq

In \cite{GK} it was argued that the wave functions
of $\theta_2, \phi_2$, which arise though
the collective coordinate quantization of
the D3-branes wrapped over the 3-cycle $(\psi,\theta_1,\phi_1)$, correspond to
eigenstates of a charged particle on $S^2$ in the presence of a 
charge $N$ magnetic monopole. Taking the gauge potential 
$A_\phi =N (1+\cos \theta)/2, \ A_\theta=0$ we find that 
%The angular factor is recognized as a wave function on $S^2$ corresponding
the ground state wave function $\sim \sin^N (\theta_2/2)$ carries 
the $J=N/2, m=-N/2$ quantum numbers.\footnote{In
a different gauge this wave function would acquire a phase. In the string
calculation it comes from the purely imaginary Chern-Simons term in
the Euclidean D3-brane action.} 
These are the $SU(2)$ quantum numbers of ${\rm det} B_2$.
Therefore, the angular dependence of $e^{-S}$ identifies ${\rm det} B_2$
as the only operator that acquires a VEV, in agreement with the gauge theory. 

The power of $r_c$ indicates that the operator dimension is $\Delta=3N/4$,
which again corresponds to the baryonic operators.
The VEV depends on the parameter $u$
as $\sim u^{3N/4}$. This is not the same as the classical scaling that would give
${\rm det} B_2= u^N $. The classical scaling is not obeyed because this is an
interacting theory where the baryonic operator acquires an anomalous dimension.

The string theoretic arguments presented in this section provide
nice consistency checks on the picture developed in this paper, and also
confirm that the Eucldean 3-brane can be used to calculate the baryonic condensate.

\section{$B$-field on the Resolved Conifold}
\label{sec-Bfield}
  Our warped resolved conifold solution written with no 
NS-NS $B$ field corresponds to a special isolated point in the space 
of gauge coupling constants. From \cite{HKO1},
the relation between coupling constants and the SUGRA background is known to be,
\beqa
 \frac{4 \pi^2}{g_1^2} + \frac{4 \pi^2}{g_2^2} &=& \frac{\pi}{g_s e^{\Phi}} \\
 \frac{4 \pi^2}{g_1^2} - \frac{4 \pi^2}{g_2^2} &=& \frac{1}{g_s e^{\Phi}} 
\left(\frac{1}{2\pi \alpha'} \int_{S^2} B_2 - \pi \right)
\eeqa
where $\Phi$ is the dilaton. Hence when $B=0$, $g_1$ is infinite.

Since the resolved conifold has a topologically non-trivial two cycle and we could turn on a 
$B$-field proportional to the volume of this cycle \cite{KW}:
\beqa
B_2 \sim \sin \theta_2 d\theta_2 \wedge d\phi_2.
\eeqa
Such a $B$-field would have no flux, $H=dB_2=0$, while still being non-trivial ($\int_{S^2} B_2 \neq 0$). 
Since there is no flux, the rest of the SUGRA solution remains 
untouched and we have a description of the gauge theory at generic coupling.

When the resolved conifold is warped by a stack of branes as we have
in this paper, the argument of \cite{KW2} continues to hold. A new
$AdS_5 \times S^5$ throat branches out at the point where the stack
is placed. This modifies the topology by introducing a new
non-trivial 5-cycle. However, the earlier two-cycle is untouched and
does not become topologically trivial. One way to see this is to
note that the new 5-cycle was the trivial cycle that could shrink to
a point at the place where the stack is placed. But the finite two
cycle of the resolution is topologically distinct from the cycles
that shrink here and hence it obviously survives the creation of a
new 5-cycle. Hence the fluxless NS-NS $B_2$ field above that naturally
exists on such a space can be used to describe the gauge theory at
generic coupling.

Had we considered a stack of D3-branes on the deformed conifold, the
situation would have been quite different, as emphasized in
\cite{KW2}. In that case, a fluxless $B_2$ field cannot be turned
on; therefore, there is no simple $SU(N)\times SU(N)$ gauge theory
interpretation for backgrounds of the form (\ref{warpedresolved})
with $ds_6^2$ being the deformed conifold metric, and $H$ the
Green's function of the scalar Laplacian on it. Of course, the
deformed conifold with a different warp factor created by self-dual
3-form fluxes corresponds to the cascading $SU(kM)\times SU(k(M+1))$
gauge theory \cite{KS,KT}.

\section{Conclusions}
\label{sec-conc}
 We have constructed the SUGRA duals of the $SU(N)\times SU(N)$ conifold gauge theory with
 certain VEV's for the bi-fundamental fields. As discussed in \cite{KW2},
 the different vacua of the theory correspond to D3-branes localized on
  the singular as well as resolved conifold.
  Vacua with ${\cal U} = 0 $ describe the singular conifold with a localized stack of
  D3-branes;
  vacua with ${\cal U} \neq 0$ instead describe D3-branes localized on the conifold resolved
  through blowing up of a 2-sphere.
We constructed explicit SUGRA solutions corresponding to
  these vacua.
In particular, the
solution corresponding to
giving a VEV to only one of the fields in the gauge theory,
$B_2 =u 1_{N\times N}$, while keeping
$ A_i = B_1 = 0$,
corresponds to 
a certain warped resolved conifold. In this case the warp factor
is given by the Green's function with a source at a point on the
blown-up 2-sphere at $r=0$. The baryonic operator $\det B_2$ gets
a VEV while no chiral mesonic operator does. This background is
thus dual to a non-mesonic, or baryonic, branch of the CFT.
To confirm this, we used the action of a Euclidean D3-brane 
wrapping a 4-cycle in the resolved
conifold, to calculate the VEV of the baryonic operator. 

The explicit SUGRA solution
was determined and found to asymptote to $AdS_5 \times T^{1,1}$ in
the large $r$ region. When one approaches the blown-up 2-sphere, the
warp factor causes an $AdS_5 \times S^5$ throat to branch off at a
point on the 2-sphere. Our calculation makes use of the explicit
metric on the resolved conifold found in \cite{PandoZayas}. Our
warped solution, with a localized stack of D3-branes, is completely
non-singular in contrast to the smeared-brane solution obtained in
\cite{PandoZayas}.

  The Green's functions on the singular and resolved conifolds were determined
in detail
  for the purpose of constructing the SUGRA solutions.
These Green's functions are also useful in
  brane models of inflation where they play a role in computing the one-loop corrections to
  non-perturbative superpotentials (see \cite{Giddings,BDKMMM} for such an application).
  The Green's functions were written using harmonics on $T^{1,1}$ in the $a,b$ variables on the conifold
  (instead of the usual angular variables or the $z,w$ co-ordinates).
  This facilitated the comparison with the explicit gauge theory operators that acquire
  VEVs. 

   We see a number of possible extensions of our work. One of them deals with
the AdS/CFT dualities
based on the Sasaki-Einstein spaces $Y^{p,q}$ \cite{Gaunt1,Hanany}.
Calculations similar to ours can be performed for the resolved cones over $Y^{p,q}$ manifolds
   (for recent work, see \cite{ChenLuPope,Oota}). Harmonics in convenient co-ordinates similar to the ones constructed
   here could perhaps be constructed using the bifundamental fields of these quiver gauge
   theories. Again, the basic
non-singular solutions will correspond to
   a stack of branes at a point, and it would be interesting to
   solve for the corresponding warp factors. One could also study
   the resolved cone versions of the solutions found in
   \cite{HEK}, which correspond to cascading gauge theories.
   It is also possible to consider Calabi-Yau cones with blown-up
4-cycles \cite{PTseyt,Pal,Sfetsos,Hanany,Benini}.
In \cite{Benvenuti}, the gauge theory operator whose VEV
corresponds to blown-up 4-cycles of certain cones was identified.
Perhaps the Green's function could be determined for a stack of branes on such 4-cycles,
   giving the non-singular SUGRA dual of corresponding non-mesonic branches in the gauge theory.

\section*{Acknowledgements}
We are indebted to Edward Witten for stimulating discussions that
suggested the way to calculate the baryonic condensate.
We also thank Sergio Benvenuti, David Kutasov, Dmitry Malyshev, Anand
Murugan, Manuela Kulaxizi, Yuji Tachikawa, Justin Vazquez-Poritz 
for useful discussions, and especially Arkady Tseytlin
for discussions and comments on the manuscript. We are also grateful
to Daniel Baumann, Marcus Benna, Anatoly Dymarsky, Juan Maldacena and Liam
McAllister for collaboration on related work, and to Daniel Baumann
for his help in making the figures. This research was supported in
part by the National Science Foundation under Grant No. PHY-0243680.
Any opinions, findings, and conclusions or recommendations expressed
in this material are those of the authors and do not necessarily
reflect the views of the National Science Foundation.

\appendix

\section{Eigenfunctions of the Scalar Laplacian on $T^{1,1}$}
\label{sec-eigen}
The main emphasis of this Appendix is on writing the harmonics on
$T^{1,1}$ in a way that makes the connection with the dual gauge
theory operators most transparent. The eigenfunctions of the scalar
Laplacian on $T^{1,1}$ have been worked out in \cite{Gubser, Ceresole}. We first review this
calculation and present the harmonics in angular variables on
$T^{1,1}$. This form of the harmonics is useful for some purposes,
such as in \cite{BDKMMM} where it was used to find the potential
generated for a D3-brane moving on the conifold due to a wrapped D7.
We then write the harmonics using the complex $a_i,b_j$ coordinates, generalizing the $z_i$ construction of \cite{KW2}, that makes the connection with the gauge theory manifest. We also construct the operators using $A_i,B_j$ with given symmetry charges, related to the harmonics through the AdS/CFT correspondance.

Since $T^{1,1}$ is a product of two $3-$spheres divided by a $U(1)$,
the eigenfunctions are simply products of harmonics on two
$3-$spheres, restricted by the fact that the two spheres share an
angle $\psi$. The laplacian (defined by $\Delta_Z H =
\frac{1}{\sqrt{g}} \partial_m (g^{mn} \sqrt{g}
\partial_n H )$) on $T^{1,1}$ can be written in the following form,
\begin{equation}
{\label{eqA1} \Delta_Z = 6 \Delta_1 + 6 \Delta_2 + 9 \Delta_R }
\end{equation}
where
\begin{equation}
\displaystyle{
\Delta_i = \frac{1}{\sin\theta_i} \partial_{\theta_i}  \;
               (\sin \theta_i \; \partial_{\theta_i} \;\;) +
               \left( \frac{1}{\sin\theta_i} \partial_{\phi_i} - \cot\theta_i
\partial_\psi \right)^2
}
\end{equation}
\begin{equation}
\displaystyle{ \Delta_R = \partial^2_\psi }
\end{equation}
We can solve for the eigenfunctions through separation of variables,
$$\displaystyle{
Y_I(Z) \sim J_{l_1,m_1,R}(\theta_1) \; J_{l_2,m_2,R}(\theta_2) \; e^{i m_1 \phi_1 + i m_2
\phi_2} \; e^{\frac{i R \psi}{2}}   }
$$
This leads to
\begin{equation}
\displaystyle{
\frac{1}{\sin\theta} \partial_{\theta}  \; (\sin \theta \;
\partial_{\theta} \; J_{lmR}(\theta)) -
               \left( \frac{1}{\sin\theta} m - \cot\theta \frac{R}{2}
\right)^2  J_{lmR}(\theta) = - E  J_{lmR}(\theta)
}
\end{equation}
for both sets of angles. When $R=0$, this reduces to the equation for harmonics
on $S^2$. For general integer $R$, this is closely related to the harmonic
equation on $S^3$ in Euler angles $(\theta, \phi, \psi)$. The eigenvalues $E$
are $l(l+1) - \frac{R^2}{4}$ as can be seen by comparing with Laplace's
equation on $S^3$.

The solutions for $J_{lmR}$ are,

\begin{equation}
\displaystyle{
J_{lmR}^{A}(\theta) =  \sin^m \theta \; \; \cot^{\frac{R}{2}}
\frac{\theta}{2} \;\; {}_2F_1 \left(-l +m, 1+l+m; 1+m -\frac{R}{2}; \sin^2\frac{\theta}{2} \right)
}\end{equation}

\begin{equation}
\displaystyle{
J_{lmR}^{B}(\theta) = \sin^{\frac{R}{2}}\theta  \;\; \cot^m
\frac{\theta}{2} \;\; {}_2F_1 \left(-l +\frac{R}{2}, 1+l+\frac{R}{2}; 1-m
+\frac{R}{2}; \sin^2\frac{\theta}{2} \right)
}\end{equation}

 Here ${}_2F_1$ is the hypergeometric function. If $m \leq R/2$, solution B is non-singular. If $m \geq R/2$, solution
A is non-singular. (The solutions coincide when $m = R/2$).

Putting together these solutions, the spectrum is of the form $$ E_I = 6 \left( l_1(l_1+1) + l_2(l_2+1) - \frac{R^2}{8} \right) $$
with eigenfunctions that transform under $SU(2)_A \times SU(2)_B$ as the spin $(l_1,l_2)$
representation and under the shift of $\psi/2$ (which is $U(1)_R$ in the UV) with charge $R$.
 Here $I$ is a multi-index with the data:
$$
I \; \equiv \; (l_1,m_1), \; (l_2,m_2), \: R
$$
with the following restrictions coming from existence of single valued regular solutions:
\begin{itemize}
\item[-] $l_1$  and $l_2$ both integers or both half-integers
\item[-] $ R \in \mathbf{Z}$  with $ \frac{R}{2} \in \{-l_1, \cdots,
l_1\} $ and $\frac{R}{2} \in \{-l_2, \cdots, l_2\}$
\item[-] $ m_1 \in \{-l_1, \cdots, l_1\}$ and $ m_2 \in \{-l_2, \cdots,
l_2\}$
\end{itemize}
As above $(l_1,l_2),R$ are the $SU(2)\times SU(2)$ spins and R-charge and
$(m_1,m_2)$, the $J_z$ values under the two $SU(2)$s.

\subsection*{Harmonics in the $a,b$ basis}

In \cite{KW2}, the 'chiral' harmonics were constructed using the complex $z_i$ coordinates. We generalize this to construct harmonics by using the $a_i,b_j$ coordinates which facilitates the comparison with the gauge . We form the eigenfunction $Y_I$ in the $a,b$ basis by tensoring
representations. As we wish to construct
harmonics on the base $T^{1,1}$, we fix the radius $r$ of the conetheory
by setting $ |a_1|^2+|a_2|^2=|b_1|^2+|b_2|^2=1$.
Since we are dealing with commuting functions (or
symmetric tensors), only the highest total spin survives the tensor
product. First we introduce the products, \beqa
\label{symmetricbricks}
 \sqrt{\frac{n!}{(2m)! (n - 2m)!}} a_1^{\frac{n}{2} + m}  a_2^{\frac{n}{2}-m} \;\; \equiv \;\;
   \vert \frac{n}{2}\;,\;\; m \rangle \qquad \left( n \in \mathbf{Z}, m \in \mathbf{Z} - \frac{n}{2} \right) \nonumber \\
\sqrt{\frac{n!}{(2m)! (n - 2m)!}}  \;\; \bar{a}_2^{\frac{n}{2} + m}
\bar{a}_1^{\frac{n}{2}-m} \;\; \equiv \;\;
   \overline{\vert \frac{n}{2}\;,\;\; m \rangle} \qquad \left( n \in \mathbf{Z}, m \in \mathbf{Z} - \frac{n}{2} \right)
\eeqa which are states of definite $SU(2)$ spin $n$ and $R$ charge
$\pm n/2$,
 since the product of $n$ commuting $a$'s and $\bar{a}$'s automatically has only
spin $n/2$ states.
%The combinatorial factors are for normalization.
We combine these to form a state of arbitrary $SU(2)$ spin and $R$
charge using Clebsch-Gordon coefficients, \footnote{We are only
using the `top-spin' Clebsch Gordon coefficients. The notation here
is:
$$
{}_RC^{l_1, m_1}_{  k; \tilde{k} } = \langle l_1,m_1 \vert
\frac{l_1}{2}+ \frac{R}{4} , k \; ; \; \frac{l_1}{2} -\frac{R}{4},
\tilde{k} \rangle \times (-1)^{\frac{l_1}{2} -\frac{R}{4}-
\tilde{k}}
$$ We need this extra $-1$ factor because we tensoring conjugate representations of $SU(2)$ : $J_- a_1 \sim a_2$ but $J_-\bar{a_2} \sim - \bar{a}_1$ }
 by,
\begin{eqnarray}
\label{wvfdefn}
 \vert l_1 , m_1;R/2 \rangle_a & =& \sum_{k,\tilde{k} \atop k + \tilde{k} =
m_1} {}_RC^{l_1, m_1}_{  k \; ;\; \tilde{k} } \vert \frac{l_1}{2}+
\frac{R}{4} , k \rangle \overline{\vert \frac{l_1}{2} -\frac{R}{4},
\tilde{k} \rangle} \nonumber \\
&= & {\displaystyle \left(a_1 a_2\right)^{\frac{l_1}{2}
+\frac{R}{4}} \left(\bar{a}_1\bar{a}_2\right)^{\frac{l_1}{2}
-\frac{R}{4}}} \sum_{k + \tilde{k} = m_1} {}_RC^{l_1, m_1}_{ k \;
;\; \tilde{k} } \;\; a_1^k a_2^{-k} \bar{a}_2^{\tilde{k}}
\bar{a}_1^{-\tilde{k}}
\end{eqnarray}
where we have introduced $\vert l_1 , m_1;R/2 \rangle_a $ to denote
the wavefunctions with $SU(2)$ spin $(l_1,m_1)$ and $U(1)_R$ charge
$R/2$ constructed from $a_i$ variables.

Using the same notation for $b_i$, $\vert l_2 , m_2;R/2\rangle_b$ is
the state with the required symmetry charges. To construct an
eigenfunction $Y_I$ on $T^{1,1}$, we must have equal $R$ charge for
the $a$ and $b$ states above in order to have invariance under the
transformation $a \to e^{i\alpha} a, b \to e^{-i \alpha} b$
explained earlier (see (\ref{removeredun})). Hence, $Y_I$ is simply
a product of the $a$ and $b$ states constructed above, \beqa
\label{compwvf}
 &Y_I& \sim \vert l_1 , m_1;R/2 \rangle_a  \vert l_2 , m_2;R/2 \rangle_b
\ .\eeqa
For example, some of the wavefunctions for $l_1=l_2=1, R=0 $ are :
\begin{eqnarray}
 \;  & a_1\bar{a}_2 \; b_1\bar{b}_2  & (m_1,m_2) = (1,1) \nonumber \\
 \;  &(a_1\bar{a}_1-a_2\bar{a}_2) \; b_1\bar{b}_2 & (m_1,m_2) = (0,1) \nonumber \\
 \; &a_2\bar{a}_1 \; b_2\bar{b}_1 & (m_1,m_2) = (-1,-1) \nonumber
\end{eqnarray}

While the harmonics (\ref{compwvf}) are obviously relevant to the
singular conifold, it was also shown in Section \ref{sec-resolved}
that the Laplacian on the resolved conifold (see (\ref{reseq2}))
factors in a form that allows one to use the same angular functions.
This is because the resolution of the conifold preserves the
$SU(2)_L\times SU(2)_R\times U(1)_R$ symmetry.

\subsection*{Construction of the dual operators }

The above construction of eigenfunctions is useful primarily because
of their one-to-one correspondence with (single trace) operators in
the guage theory. Our stragey is to replace $a_i,b_j$ in the
eigenfunctions by the chiral superfields $A_i,B_j$. However, since
$A_i,B_j$ are non-commuting operators in the gauge theory, we need
to modify the procedure of the previous section to obtain an
operator ${\cal O}_I$ of a given symmetry.

We may start with (\ref{symmetricbricks}), and symmetrize the
product of $A_1, A_2$'s (and $\bar{A}_1, \bar{A}_2$'s) by hand (the
gauge index structure seems ill defined but this will be fixed when
the total operator is put together.) So we could now write instead
of (\ref{symmetricbricks}) (with a different normalization factor),
\beqa
   \frac{1}{\sqrt{\frac{n!}{(2m)! (n - 2m)!}}}  \sum_{\frac{n}{2} + m = \sum i \atop \frac{n}{2}-m = \sum j} A_1^{i_1}  A_2^{j_1} A_1^{i_2} \cdots A_2^{j_k} \;\; \equiv \;\;
   \vert \frac{n}{2}\;,\;\; m \rangle \qquad \left( n \in \mathbf{Z}, m \in \mathbf{Z} - \frac{n}{2} \right)
\eeqa
 The same symmetrization applies to $\bar{A}$'s as well.
With this modified definition of $\vert \frac{n}{2},m \rangle$, we
can write down the equation analogous to (\ref{wvfdefn}) with no
change in the form,
 \beqa
 \vert l_1 , m_1;R/2 \rangle_A & =& \sum_{k,\tilde{k} \atop k + \tilde{k} = m_1} {}_RC^{l_1, m_1}_{  k \; ;\; \tilde{k} } \vert \frac{l_1}{2}+ \frac{R}{4} , k \rangle \overline{\vert \frac{l_1}{2} -\frac{R}{4} , \tilde{k} \rangle}
 \eeqa

 We make the analogous definitions for $B$.
Finally, we can write down dual operator ${\cal O}_I$ as,
 \beq \label{operatorL}
\displaystyle {\cal O}_I =  {\rm Tr} \left(\vert l_1 , m_1;R/2
\rangle_A  \vert l_2 , m_2;R/2 \rangle_B\right) \eeq

 The product of the operators $\vert l_1 , m_1;R/2 \rangle_A$ and $\vert l_2 ,
m_2;R/2 \rangle_B$ is taken in the following way. All the terms are
multiplied out and in each term, one is free to move operators in
the $(N,\bar{N})$ rep of the gauge group (i.e $A,\bar{B}$) past
$(\bar{N},N)$ (i.e $B,\bar{A}$) but no rearrangement among
themselves is allowed. We shuffle them past each other until they
alternate and so we can contract gauge indices properly and take the
trace. It is easy to verify that the numbers of fields of each type
are equal and so there is always one essentially unique way of doing
this. By construction, this operator has the specified symmetry $I$
under the global symmetry group.

\section{$AdS_5 \times S^5$ Throats in the IR}
\label{sec-throat}
Here we show explicitly that the Green's function on the resolved conifold
reduces to the form $\frac{1}{y^4}$ near the source as it must
($y$ here is the physical distance from the source
on the transverse space).
This leads to the usual near-horizon limit when the branes are at a smooth point and
hence an $AdS_5 \times S^5$ throat. This is of course to be expected since close to the source,
we can find coordinates in which the space looks flat at leading order and
hence the Green's function must behave as $\frac{1}{y^4}$. But it is instructive to see
how the series does add up to such a divergence while each individual term has a
different kind of divergence that gives a singular geometry 
in the case of the resolved conifold.

We focus on the resolved conifold and consider $\theta_0 = 0 = \phi_0$, i.e set the stack on
the 'north pole' of the finite $S^2$. Also, we approach the singularity by first setting
$\theta_2 = 0$ and taking the $r \rightarrow 0$ limit. 
Now $r$ is the physical distance and from (\ref{resolvedgreenN}), we would like to show that,
\beq \label{RTP}
\sum_{l} (2 l + 1) H_I^A(r) \sim \frac{1}{r^4}  \quad \mbox{ while } 
\quad  H_I^A(r) \sim \frac{1}{r^2} \quad \mbox{ as } r \rightarrow 0
\eeq

Consider the regulation of the sum of squares of integers,
\beq
\sum_{n=0}^{\infty} n^2 \to   \sum_{n=0}^{\infty} n^2 R(n \epsilon)
\eeq
where $R(x)$ is a regulator such as $R(x) = e^{-x}$ with the property $R(x) \rightarrow 0$ (fast enough in a sense to be seen below) as $x \rightarrow \infty$. As $\epsilon \to 0$, the sum diverges and this allows one to approximate the sum by an integral in this limit. Further, only $0 \leq n \leq 1/\epsilon$ will contribute. Hence we find,
\beq \label{dimensional}
 \int_0^{\frac{1}{\epsilon}} n^2 R(n \epsilon) dn \;\;\; \sim \;\;\; \frac{1}{\epsilon^3} \int_0^{1} y^2 R(y) dy \quad (\epsilon \to 0)
\eeq

Note that the above argument just amounts to dimensional analysis.
To cast the given expression (\ref{RTP}) in the above form with $
H_I^A(r)$ playing the role of a regulator, we note that $H_I^A(r)$ can be approximated for
$r \ll a$ by $(\sqrt{l(l+1)}/r) K_1(\sqrt{l(l+1)}r)$.
\footnote{We mean this in the sense that $K_1(y)$ is the solution to the differential
equation obtained by applying $r \ll a$ to (\ref{resradialeqn}) whose exact solution
was obtained as $H_I^A(r)$. We are interested in how $r$
scales with $l$ to keep $H_I^A(r)$ constant for very small $r$,
since this determines the leading order singularity through essentially 
dimensional analysis in
(\ref{dimensional}).
Approximating $H_I^A$ by $K_1$ is valid in this sense. } Hence we have for $r \ll a$,
\beqa
\sum_{l} (2 l + 1) H_I^A(r) & \sim &  \sum_{l} (2l+1)\frac{\sqrt{l(l+1)}}{r} 
K_1(\sqrt{l(l+1)} r) \sim \frac{1}{r} \int_n (2n+1)n K_1(n r) \\
& \sim & \frac{1}{r} \int_0^{1/r} n^2 K_1(n r) dn \sim \frac{1}{r} \times \frac{1}{r^3} \;\;\; \int_0^{1} y^2 K_1(y) dy
\eeqa
where we have kept track of only the leading order singularity. We have identified 
$R(y) = K_1(y)$ despite the fact $K_1(y) \sim 1/y$ for small y. 
This is allowed here because $\int_0^1 dy y^2 K_1(y)$ converges.

Hence we see that indeed, $H(r,\theta_2=0) \sim \frac{1}{r^4}$ near $r=0$ and hence the geometry is non-singular (though each term in the expansion of $H$ behaves as $\frac{1}{r^2}$ giving a singular geometry by itself). The result essentially follows from dimensional analysis in (\ref{dimensional}).

\bibliographystyle{unsrt}

\end{document}